\documentclass{article}

\begin{document}

\title
{Some Effects of Classical Feedback on the Classical Capacity of a
Memoryless Quantum Channel}

\author{Gleb V. Klimovitch \\
Information System Laboratory \\
Stanford University \\
Stanford, CA\\
\tt{gleb@stanford.edu}}

\maketitle
\begin{abstract}
Classical feedback is defined here as the knowledge by the
transmitter of the quantum state of the qubit received by the
receiver. Such classical feedback doubles capacities of certain
memoryless quantum channels without preexisting entanglement
between transmitter and receiver. The increase in capacity, which
is absent on classical memoryless channels, occurs because we can
transform an entangled qubit pair into any other entangled state
by applying a unitary operator to only one of the qubits.
\end{abstract}

\section{Introduction}
For a classical channel, feedback can be naturally defined as the
knowledge of received symbol at the transmitter after transmission
\cite{StdDefFB}. Quantum case is more subtle, since physical
arrival of a qubit at the receiver does not necessarily mean that
the receiver knows the state of the received qubit and/or can copy
the unknown state to the transmitter. In this paper, we still use
the definition of classical feedback given above for a quantum
channel. In particular, we assume that the received state can be
known to the transmitter in the presence of channel noise, before
it is measured by and therefore known to the receiver. Such an
assumption may appear counterintuitive at first glance, but it
does work for some channels.

Consider the following (fancy but workable) example. The channel
either flips the spin of the qubit or does not affect the qubit at
all. There are extra "helper qubits" in the channel that are
represented by particles different from "information qubits". The
helper qubits are independent identically distributed, each of
them being in only two orthogonal states, e.g. spin-up and
spin-down, with equal probabilities one half. The helper qubits
are neither sent nor affected by the transmitter, but the
transmitter learns the initial state of each helper qubit, e.g. by
receiving its "carbon copy". Both helper qubits and information
qubits are equally affected by the channel. The receiver can
separate the two kinds of qubits and measure only the states of
the received helper qubits. The receiver still cannot detect spin
flips, since it does not know the initial states of the helper
qubits". The transmitter obviously can, once it gets the feedback
from the receiver.

Thus defined, classical feedback increases the classical capacity
of certain memoryless quantum channels, as demonstrated by a
simple example in Section 2. In Section 3, we argue that increase
in capacity due to feedback becomes possible for non-classical
(quantum) channels due to distributed nature of information stored
by qubits.

\section{How it works}

We consider a quantum subchannel, whose effect on qubit is
described by random unitary operators
$\widehat{U}_{channel}(\mathbf{c})$ with probability distribution
$p(\mathbf{c})$ for random vector $\mathbf{c}$.  For memoryless
channel, operators $\widehat{U}_{channel}$ corresponding to
different transmission events are statistically independent.

Our quantum channel consists of two quantum subchannels with
different noise levels. Let each channel transmit one qubit per
unit of time. The main idea is best illustrated by assuming that
one subchannel is so noisy that its capacity approaches zero,
while the other subchannel is noiseless.

For example let the noisy subchannel be disturbed by random
magnetic field which interacts with magnetic moment of qubits, so
that qubit states are completely randomized (but not collapsed)
during transmission. The effect of the noisy subchannel on the
qubit can be modeled by random unitary operators
\begin{equation}
\label{Uchannel} \widehat{U}_{channel}(\mathbf{c})=exp(j\lambda
\mathbf{c}\widehat{\mathbf{\sigma}})\;,
\end{equation}
where vector $\widehat{\sigma}$ has Pauli matrices as components;
$\mathbf{c}$ is a random three-dimensional gaussian vector with
zero mean and unit variance per dimension; and $\lambda\gg 1$ is a
constant proportional to the product of a typical channel field,
the magnetic moment of the qubit, and transmission time.

Without feedback, the capacity of the noisy subchannel approaches
zero, as $\lambda$ goes to infinity. Information can be sent only
over the quiet subchannel with the maximum load of one bit per
transmission (without preexisting entanglement between transmitter
and receiver). Thus the channel capacity equals one - the rate of
qubit trasmission over each subchannel.

Feedback doubles the capacity as follows. Transmitter forms
entangled pairs of qubits and sends members of each pair at
different times. First, one member of the pair goes over the noisy
subchannel. After it is received, the transmitter learns the
channel estimate during its transmission (let it be vector
$\mathbf{c}$). Then the trasmitter undoes the effect of the noisy
subchannel on the quantum state of the pair by applying a unitary
operator $\widehat{U}_{undo}(\mathbf{c})$ to the second member of
the pair (that is still at the transmitter).

For example, if the pair is in the singlet state, then
$\widehat{U}_{undo}(\mathbf{c})$ is also given by the r.h.s. of
equation (\ref{Uchannel}) but is applied to the second rather than
the first member of the pair. Indeed, the combined action of
channel and "undo" operators on the pair can be rewritten in terms
of the (operator of the) total spin of the pair
$\widehat{\mathbf{S}}$ as follows:
\begin{equation}
\widehat{U}_{undo}\widehat{U}_{channel}=exp(2j\lambda
\mathbf{c}\widehat{\mathbf{S}})=\widehat{1}\;.
\end{equation}
The last equality holds because the spin of the pair is zero.

After the action of the noisy subchannel on the pair is undone,
the transmitter applies the superdense quantum coding
\cite{SuperDenseCoding} of two bits per qubit to the second member
of the pair and transmits it over the quiet subchannel
(simultaneously with transmitting the first member of another pair
over the noisy subchannel). The receiver waits until both members
of the pair arrive, then the pair is decoded. Thus the channel
capacity doubles to two - the combined rate of qubit trasmission
over both subchannels. Despite one subchannel being extremely
noisy, we achieve the maximum possible capacity (without
preexisting entanglement between transmitter and receiver) of one
bit per qubit due to feedback.

\section{Why it works}
Let us start with an insight "why it does not work" for classical
channels, i.e. why feedback does not increase the capacity of a
classical memoryless channel. The feedback gives the trasmitter
the knowledge of decoder errors. However, transmission at a higher
rate but with errors necessitates transmission of information to
correct such errors, which in turn reduces the number of bits
available to send new data. As a result, the capacity does not
increase due to feedback on a classical memoryless channel. For
example, a classical analog of our system consists of one
extremely noisy and one quiet subchannel of equal (unit) rates.
Without feedback, we would just ignore the noisy subchannel. With
feedback, we could use the quiet subchannel to correct all the
errors on the noisy subchannel. However, such an error correction
would leave no resources to send new data over the quiet
subchannel, and the capacity remains the same.

In quantum case, information  distributed between two entangled
qubits as follows. While both qubits are needed to decode
anything, it suffices to access one qubit to encode (or re-encode)
two bits, i.e. the maximum amount of information per qubit pair
(unless the pair itself is entangled with the receiver). As a
result, even after sending the first qubit in the pair and having
only the second qubit at its disposal, the trasmitter can still
cancel the effects of the noisy subchannel on the whole pair and
re-encode the whole pair. Speaking informally, it is the miracle
of superdense quantum coding \cite{SuperDenseCoding} that helps to
increase the capacity of such a quantum channel by feedback.

\section{Summary}
Defined as the knowledge of already received state at the
trasmitter, classical feedback on memoryless quantum channel can
increase (double) the channel capacity. The increase is achieved
by a method similar to superdense quantum coding.

\section{Acknowledgements}
The author strongly appreciates discussions with Prof. Thomas
Cover on topics in both classical and quantum information theory.

The author will be quite grateful for feedback on this paper,
including comments on and references to related work, and hopes
that such a feedback could increase the "channel capacity" of the
paper.

\end{document}